# SPLITTING OF THE ALFVÈN SURFACE IN A RELATIVISTIC PULSAR WIND

H. Ardavan

*Institute of Astronomy, University of Cambridge,
Madingley Road, Cambridge CB3 0HA, UK*

**Abstract**

In a recent paper, Li and Melrose have claimed that the splitting—due to relativistic effects—of the Alfvèn surface in an axisymmetric pulsar wind does not occur. Here we refute this claim by showing that, unless the solution that describes the flow along each open magnetic field line passes through the pure Alfvènic point (which is one of the manifestations of the splitting of the Alfvènic point), it would not be physically viable both at the surface of the star and at infinity.

## 1. Introduction

Under the conditions of axial symmetry and time-independence, several first integrals of the system of magneto-hydrodynamic equations that govern a relativistic pulsar wind are known (Ardavan 1976). The conservation laws for mass, energy and angular momentum, together with Ferraro's isorotation law, enable us to establish a relationship between the values of any two flow variables along a given open magnetic field line. Each of these relationships can generally be written in the form of an algebraic equation for one variable as a function of the other. For instance, the Lorentz factor of the plasma satisfies a quadratic equation whose coefficients depend on the plasma density and the distance of the field point from the axis of symmetry. Amongst the various solutions of such an algebraic equation, there is normally none that can describe the entire flow on its own: the solution that satisfies the required boundary conditions at the surface of the star fails to have an acceptable asymptotic behaviour at infinity, and the solution that behaves properly at infinity becomes unphysical at the star.

A physically acceptable description of the flow can therefore be obtained only by matching two (or more) solutions of the algebraic equation in question along a given magnetic field line. If the flow is to be free of discontinuities, however, these distinct solutions must be matched across surfaces at which the discriminant of the equation vanishes without changing sign. Such zeros of the discriminant of each equation define the loci of the critical (or singular) points through which the composite solution describing the actual flow should pass. The Alfvènic singularity—at which the poloidal component of the flow velocity equals the local propagation speed of Alfvèn waves—is one of these points.

Of the several critical points thus arising from the various algebraic equations that are implied by the first integrals of the equations of motion, there are two—the so-called pure Alfvènic and intermediate critical points—which coalesce and coincide with the Alfvènic point in the non-relativistic regime. Not only these offshoots of the Alfvènic point, but every one of the critical points that arise in the context of constructing physically acceptable composite solutions must necessarily occur within the flow.

There are other contexts in which the Alfvènic point and the manifestations of its relativistic splitting arise. From the first integrals of the equations of motion, one can also derive expressions for certain flow variables which have the form of a ratio. It so happens that the zeros of the denominators in such ratios often coincide with the critical points of the flow as identified above. Li & Melrose (1994) define the critical points or singularities of the flow by means of such ratios alone and, from the fact that the numerators and the denominators in these ratios have common factors when expressed in terms of certain variables, draw the conclusion that the pure Alfvènic and the intermediate singularities are spurious.

If the only argument for the occurrence of these critical points were the possibility of the vanishing of the denominators in the expressions that one obtains for various flow variables, then the observations made by Li and Melrose would have implied that this argument does not suffice to establish the occurrence of the pure Alfvènic and the intermediate critical points: the zeros of the numerators and the denominators in the expressions in question may automatically coincide. However, as we shall see in §3, the main argument for the splitting of the Alfvèn surface is based on the global behaviour of the solutions of certain quadratic equations and is in fact totally independent of the ratios by means of which Li and Melrose define the singularities of the flow.

## 2. Formulation of the problem: the Alfvèn surface

The magneto-hydrodynamic equations which govern a relativistic pulsar wind have the following first integrals when the flow is time-independent and axially symmetric:

$$\mathbf{v} = \kappa \mathbf{B} + r\omega \hat{\mathbf{e}}_\varphi, \tag{1}$$

$$\kappa \rho = F, \tag{2}$$

$$-rB_\varphi/(4\pi) + r\gamma\rho v_\varphi \kappa = G, \tag{3}$$

$$\gamma\left(1 - \frac{r\omega}{c}\frac{v_\varphi}{c}\right) = \exp H \tag{4}$$

(see Ardavan 1976). In these expressions, $\mathbf{v}$ is the flow velocity, $\mathbf{B}$ is the magnetic field, $\rho$ is the rest-mass density,

$$\gamma = (1 - v^2/c^2)^{-\frac{1}{2}} \tag{5}$$

is the Lorentz factor, $\kappa$ is a scalar function of position, $r$ is the distance from the axis of symmetry, $\hat{\mathbf{e}}_\varphi$ is the unit vector associated with the azimuthal direction and the subscript $\varphi$ designates the toroidal component of a vector. The quantities $\omega, F, G$, and $H$ remain constant along the magnetic lines of force. If we exclude the possibility of the differential rotation of the magnetic field lines, then $\omega$ represents the constant value of the angular velocity of rotation of the rigid surface of the central neutron star. The remaining constants $F, G$ and $H$ are related to the rates of transport of mass, angular momentum and energy per unit flux-tube, respectively.

Equations (1)–(5) comprise six scalar relationships between the values of the coordinate $r$ and the seven variables $v_p, v_\varphi, B_p, B_\varphi, \rho, \kappa$ and $\gamma$ along a given magnetic field line ($v_p$ and $B_p$ denote the poloidal components of $\mathbf{v}$ and $\mathbf{B}$). They can be employed to express six of these variables as functions of the seventh variable and $r$. In particular, one can express $\gamma$ as a function of $4\pi F\gamma\kappa = 4\pi\gamma\rho v_p^2/B_p^2$ and $\hat{r} \equiv r\omega/c$ to obtain

$$\gamma = \frac{(1 - \hat{r}_A^2 - 4\pi F\gamma\kappa)\exp H}{(1 - \hat{r}_A^2)(1 - \hat{r}^2 - 4\pi F\gamma\kappa)}, \tag{6}$$

in which

$$\hat{r}_A \equiv \left(1 + \frac{c^2 F \exp H}{\omega G}\right)^{-\frac{1}{2}}. \tag{7}$$

The covariant representation of the statement that the velocity of the plasma in the polidal direction equals the Alfvèn speed in the present case reduces to

$$4\pi\gamma\rho v_p^2/B_p^2 = 1 - \hat{r}^2 \tag{8}$$

(see the Appendix in Ardavan 1976). At the Alfvèn surface defined by (8) the denominator in (6) vanishes. Requiring that $\gamma$ should remain finite at this surface, i.e. that the numerator of the expression in (6) should vanish simultaneously with its denominator, we find that the open lines of force of the magnetic field cross the Alfvèn surface at $\hat{r} = \hat{r}_A$.

The Alfvèn cylinder (8) (whose generator is not in general a straight line) constitutes the surface of parabolic degeneracy of the system of partial differentail equations that governs the axisymmetric pulsar wind: this system, which undergoes a change in type from elliptic to hyperbolic across the magnetosonic surface $4\pi\gamma\rho v_p^2 + \hat{r}^2 B_p^2 - B^2 = 0$, becomes parabolic also at $\hat{r} = \hat{r}_A$, within its domain of ellipticity (see Ardavan 1979). So, another way of inferring the occurrence of the Alfvènic singularity is to analyze the characteristics of the governing field equations. The calculation presented in §4 of Li & Melrose (1994) is part of the linearized version of such an analysis.

The elimination of $B_\varphi$ between (1) and (3) results in an expression for $v_\varphi$ which has a structure similar to that of the expression on the right-hand side of (6):

$$v_\varphi = \frac{r\omega - 4\pi G\kappa/r}{1 - 4\pi F\gamma\kappa}. \tag{9}$$

In the non-relativistic regime, where the denominators in (6) and (9) coincide, this expression, too, becomes singular at the Alfvèn surface and has to be regularized.

In the relativistic regime, however, the surface

$$1 - 4\pi F\gamma\kappa = 0 \tag{10}$$

at which the denominator in (9) vanishes is distinct from the Alfvèn surface and the question arises as to whether this surface ever occurs within the flow. If the so-called pure Alfvènic surface that is defined by (10) exists, then at this surface we should also have

$$r\omega - 4\pi G\kappa/r = 0, \tag{11}$$

for the toroidal component of the flow velocity, $v_\varphi$, should be everywhere finite.

## 3. Occurrence of the pure Alfvènic singularity

That the flow does in fact pass through surface (10) can be seen by considering the relationship between the Lorentz factor $\gamma$ and the density $\rho$ of the plasma along an open field line. We show in this section that $\gamma$ is given as a function of $\rho$ and $r$ by a quadratic equation neither of whose solutions provides a physically acceptable description of the entire flow on its own. The solution that describes the flow close to the star is different from the one which has the correct asymptotic behaviour at infinity so that the two solutions of this quadratic equation have to be matched across the surface at which the discriminant of the equation vanishes. The surface at which the discriminant of the quadratic equation in question vanishes, on the other hand, is precisely the same as the pure Alfvènic surface that we encountered in (10).

If we use (1) and (4) to eliminate $B_\varphi$ and $v_\varphi$ from (3), and express $\kappa$ in the resulting expression in terms of $\rho$, we arrive at the following quadratic equation for $\gamma$:

$$\Gamma^2 - [(1-\hat{r}^2)R + (1-\hat{r}_A^2)^{-1}]\Gamma + R = 0, \tag{12}$$

where

$$\Gamma \equiv e^{-H}\gamma, \tag{13}$$

$$R \equiv (4\pi F^2 e^H)^{-1}\rho, \tag{14}$$

and $\hat{r}_A$ is defined in (7). Of the two solutions

$$\Gamma_\pm = \tfrac{1}{2}[(1-\hat{r}^2)R + (1-\hat{r}_A^2)^{-1}] \pm \{\tfrac{1}{4}[(1-\hat{r}^2)R + (1-\hat{r}_A^2)^{-1}]^2 - R\}^{\tfrac{1}{2}} \tag{15}$$

of this quadratic equation, only $\Gamma_-$ passes through the Alfvènic singularity: when $\hat{r} = \hat{r}_A$, it is $\Gamma_-$ that equals $(1-\hat{r}_A^2)R$ and so satisfies the Alfvènic condition (8).

The solution $\Gamma_-$, however, does not yield a positive value for the Lorentz factor at infinity: it either vanishes or becomes negative as $\hat{r} \to \infty$. Since the plasma density $\rho$—and hence $R$—reduces to zero as $\hat{r} \to \infty$, (15) yields $\lim_{\hat{r}\to\infty}\Gamma_- = 0$ if the limiting value of $-\hat{r}^2R + (1-\hat{r}_A^2)^{-1}$ is non-negative and yields $\lim_{\hat{r}\to\infty}\Gamma_- < 0$ if the limiting value of $-\hat{r}^2R + (1-\hat{r}_A^2)^{-1}$ is negative. Not only does it not yield a physically acceptable value for the Lorentz factor at infinity, but the solution $\Gamma_-$ also predicts a flux of plasma angular momentum in $\hat{r} > 1$ that is directed towards the star.

To see this, let us note that $v_\varphi$ is positive, and so the flux of plasma angular momentum is outward (cf. Ardavan 1976), only if $\Gamma > 1$ [see (4)]. In order that $\Gamma_-$ exceeds unity, i.e. that

$$(1-\hat{r}^2)R - (1-2\hat{r}_A^2)(1-\hat{r}_A^2)^{-1} > \{[(1-\hat{r}^2)R + (1-\hat{r}_A^2)^{-1}]^2 - 4R\}^{\tfrac{1}{2}}, \tag{16}$$

we need to have both

$$(1-\hat{r}^2)R - (1-2\hat{r}_A^2)(1-\hat{r}_A^2)^{-1} > 0, \tag{17}$$

and
$$\hat{r}^2 R - \hat{r}_A^2 (1 - \hat{r}_A^2)^{-1} > 0, \tag{18}$$

where the second inequality follows from squaring (16). Inequality (17) can be satisfied in $\hat{r} > 1$ only if $\frac{1}{2} < \hat{r}_A^2 < 1$. In this regime, however, it is not possible to satisfy (17) and (18) simultaneously: the upper limit set on $R$ by (17) is smaller than the lower limit on this quantity that is required by (18).

On the other hand, the solution $\Gamma_+$ which has the correct behaviour at infinity does not yield an outward flux of electromagnetic angular momentum in $\hat{r} < \hat{r}_A$. For this flux to be directed away from the star, $B_\varphi$ should be negative (see Ardavan 1976) and so, according to (1) and (4), $(1-\hat{r}^2)\Gamma$ should be smaller than unity. In order that $(1 - \hat{r}^2)\Gamma_+$ is smaller than unity, i.e. that

$$\{[(1-\hat{r}^2)R+(1-\hat{r}_A^2)^{-1}]^2 - 4R\}^{\frac{1}{2}} < 2(1-\hat{r}^2)^{-1} - (1-\hat{r}_A^2)^{-1} - (1-\hat{r}^2)R, \tag{19}$$

we need to have both

$$2(1 - \hat{r}^2)^{-1} - (1 - \hat{r}_A^2)^{-1} - (1 - \hat{r}^2)R > 0, \tag{20}$$

and

$$\hat{r} > \hat{r}_A, \tag{21}$$

where (21) follows from squaring (19). Inequality (21) cannot be satisfied in $\hat{r} < \hat{r}_A$ so that the solution $\Gamma_+$, too, is incapable of describing the entire flow on its own. This solution fails both to pass through the Alfvènic singularity and to describe the braking effect of the magnetic field in the sub-Alfvènic region.

A physically viable description of the flow thus entails the matching of $\Gamma_-$ and $\Gamma_+$ across a surface in $\hat{r} > \hat{r}_A$. The two solutions $\Gamma_-$ and $\Gamma_+$ of the quadratic equation (12) are equal to one another at points where the discriminant of this equation is zero. For the resulting composite solution to be free from discontinuities, it is therefore necessary that the two solutions match across a surface at which the discriminant of (12) vanishes, i.e. at which

$$[(1 - \hat{r}^2)R + (1 - \hat{r}_A^2)^{-1}]^2 - 4R = 0, \tag{22}$$

and (12) reduces to

$$\Gamma = \tfrac{1}{2}[(1 - \hat{r}^2)R + (1 - \hat{r}_A^2)^{-1}]. \tag{23}$$

Equations (22) and (23) jointly yield the relationship $\Gamma = R$ which defines the pure Alfvènic surface [see (10), (13) and (14)].

The position of the pure Alfvènic surface now follows from (11), (22) and (23):
$$\hat{r} = \hat{r}_A/(1 - \hat{r}_A^2)^{\frac{1}{2}}. \qquad (24)$$
Equation (24) shows that the pure Alfvènic singularity always occurs beyond the Alfvènic singularity and that at this surface $\Gamma = R = 1$ [see (22) and (23)]. It also implies that, in the ultra-relativistic or massless limit where $\hat{r}_A \to 1$ [see (7)], the pure Alfvènic singularity is relegated to infinity (cf. Okamoto 1978).

Note that the discriminant of equation (12) should vanish at the pure Alfvènic surface without changing sign, for otherwise (12) will not have real solutions on both sides of this surface. In other words, the discriminant in question and its gradient along a magnetic field line should vanish simultaneously. This requirement further constrains the flow by specifying the value of $\mathbf{B} \cdot \nabla R$ at the Alfvènic surface. It also ensures that $\Gamma_-$ and $\Gamma_+$ have finite gradients at the pure Alfvènic surface: the numerators and the denominators in the expressions for $\mathbf{B} \cdot \nabla \Gamma_\pm$ vanish simultaneously once this requirement is met [see (15)].

The above argument shows that, unless the solution that describes the flow along each open magnetic field line passes through the pure Alfvènic point, it would not be physically viable both at the surface of the star and at infinity. This argument does not, however, guarantee that the solution that passes through the pure Alfvènic point is acceptable. A physically acceptable solution needs to meet not only the constraints associated with the pure Alfvènic point but also those associated with the other critical points of the flow. It turns out that the conservation laws (1)–(4) and the equation of the mixed type that governs the poloidal magnetic field do not admit of a continuous solution which can satisfy all such constraints (Ardavan 1979, Bogovalov 1994).

# 4. Concluding remarks

The question on which the conclusions of Li & Melrose (1994) disagree with those of Okamoto (1978) and Ardavan (1979) is whether the flow passes through the pure Alfvènic singularity defined by (10) or not, i.e. whether there exists a surface within the pulsar wind at which the flow variables satisfy the relationships (10) and (11) or not. The answer to this question is independent of how one defines a singularity. Either the flow is further accelerated after passing through the Alfvènic point (8) to attain the higher value of the Lorentz factor that is required by the pure Alfvènic condition (10) or it is not. The answer to the question is dictated only by the physical conditions at the surface of the star and at infinity and by the requirements of the conservation laws (1)–(4).

What Li and Melrose observe is that the form assumed by expression (9) does not by itself establish the occurrence of the pure Alfvènic singularity because the numerator and the denominator in this expression may have a common factor. Even if the only argument for the occurrence of the pure Alfvènic point had been the possibility of the vanishing of the denominator in (9), this observation would not have ruled out the occurrence of the pure Alfvènic point as Li and Melrose claim; it would have only rendered the argument in question inconclusive. As it is, the possibility of the vanishing of the denominator in expression (9) is not in fact the only argument for the occurrence of the pure Alfvènic point.

One can establish the occurrence of the pure Alfvènic point independently of expression (9) and only use this expression to further constrain the flow. The constraint that follows from requiring the numerator in (9) to vanish simultaneously with its denominator, moreover, has as much physical significance as that which has to be satisfied at the Alfvènic point: $\Gamma = R = 1$ and (24) relate the constants of motion ($F, G, H$ and $\omega$) to the values of the flow variables at the singular point in exactly the same way as do (8) and $\hat{r} = \hat{r}_A$.

The flaw in the reasoning of Li & Melrose does not lie solely in their assuming that the splitting of the Alfvèn surface has to be inferred from the ratios they consider. They also overlook the fact that different singularities of the flow are exhibited by the relationships between different pairs of flow variables (Ardavan 1979). That the numerator and the denominator in the expression for $v_\varphi$ have common factors when expressed as functions of the coordinate $x$ ($\equiv r\omega/c$) and the flow variable $y$ ($\equiv 4\pi F^2 \gamma/\rho$) does not imply that the pure Alfvènic singularity is spurious; it only means that the relationship between the particular flow variables $v_\varphi$ and $y$ is singularity-free.

The isolated Alfvènic singularity of the non-relativistic stellar wind in the present case splits into three distinct singular points: the pulsar wind also includes an intermediate singular point, lying between the Alfvènic and the pure Alfvènic singular points, at which $4\pi\gamma\rho\kappa^2 = 1 - \hat{r}v_\varphi/c$. The analysis establishing the occurrence of this intermediate singular point entails the study of the global behaviour of the solutions of a quadratic equation which relates $\gamma\kappa$ to $\rho$. This analysis closely parallels that presented here and has already been outlined in Ardavan (1979).